\documentclass[11pt]{article}

\usepackage{latexsym}    %
\usepackage{meaning}     %
\usepackage{times}       %

\def\squareforqed{\hbox{\rlap{$\sqcap$}$\sqcup$}}
\def\qed{\ifmmode\squareforqed\else{\unskip\nobreak\hfil
\penalty50\hskip1em\null\nobreak\hfil\squareforqed
\parfillskip=0pt\finalhyphendemerits=0\endgraf}\fi}

\newenvironment{pf}{\noindent{\it Proof.\/} }{\vspace{0.1in}}

\newtheorem{prop}{Proposition}[section]
\newtheorem{defn}[prop]{Definition}
\newtheorem{lem}[prop]{Lemma}
\newtheorem{cor}[prop]{Corollary}

\newcommand{\COMMENTOUT}[1]{}

\newcommand{\indexedsupset}[2]{\ensuremath{\left\{#1^{1},\ldots,#1^{#2}\right\}}}
\newcommand{\indexedtuple}[2]{\ensuremath{\left(#1_{1},\ldots,#1_{#2}\right)}}

\newcommand{\seqset}[2]{\ensuremath{\left\{#1,\ldots,#2\right\}}}
\newcommand{\seqsetto}[1]{\seqset{1}{#1}}
\newcommand{\applied}[3]{\ensuremath{{#1}(#2^{1},\ldots,#2^{#3})}}
\newcommand{\appliedsub}[3]{\ensuremath{{#1}(#2_{1},\ldots,#2_{#3})}}
\newcommand{\appliedtuple}[4]{\ensuremath{\left(\applied{#1}{{#2}_{1}}{#3},\ldots,
  \applied{#1}{{#2}_{#4}}{#3}\right)}}

\newcommand{\bool}{\ensuremath{\mathcal{B}}}
\newcommand{\boolm}[1]{\ensuremath{\mathcal{B}^{#1}}}

\newcommand{\bottom}{\ensuremath{\bot}}
\newcommand{\true}{\mbox{\it tt}}
\newcommand{\false}{\mbox{\it ff}}

\newcommand{\function}[3]{\ensuremath{{#1}: {#2} \rightarrow {#3}}}
\newcommand{\boolfn}[2]{\function{#1}{\boolm{#2}}{\bool}}

\newcommand{\setsize}[1]{\left| {#1} \right|}
\newcommand{\miff}{\ensuremath{\Leftrightarrow}}
\newcommand{\choosefn}[2]{\ensuremath{\left(\begin{array}{c} #1 \\ #2 
                                            \end{array} \right)}}

\newcommand{\preseq}[3]{\ensuremath{S^{#1,#2}_{#3}}}
\newcommand{\preseqseq}[3]{\preseq{\{1, \ldots, #1\}}{\{1, \ldots, #2\}}{#3}}

\newcommand{\tuplein}[2]{\ensuremath{#1 \in #2}}
\newcommand{\tuplenotin}[2]{\ensuremath{#1 \not \in #2}}

\newcommand{\tuplearray}[4]{\ensuremath{\indexedtuple{{#1}^{1}}{#2}, \cdots, \indexedtuple{{#1}^{#3}}{#2} \in #4}}

\newcommand{\pareq}{\ensuremath{\equiv}}
\newcommand{\parleq}{\ensuremath{\preceq}}
\newcommand{\parnleq}{\ensuremath{\not\preceq}}

\newcommand{\bucc}[2]{\ensuremath{\mathrm{BUCC}_{(#1,#2)}}}
\newcommand{\funpor}{\ensuremath{\mathrm{POR}}}
\newcommand{\funbp}{\ensuremath{\mathrm{BP}}}
\newcommand{\fundet}{\ensuremath{\mathrm{DET}}}
\newcommand{\funtdet}{\ensuremath{\mathrm{ttDET}}}
\newcommand{\funntdet}[1]{\ensuremath{\mathrm{ttDET}_{#1}}}
\newcommand{\por}[1]{\ensuremath{\funpor_{#1}}}
\newcommand{\gust}[1]{\ensuremath{\mathrm{GUST}_{#1}}}
\newcommand{\bgust}[2]{\ensuremath{\mathrm{BGUST}_{#1}^{#2}}}
\newcommand{\bgustf}[1]{\bgust{#1}{}}

\newcommand{\cc}{\mathop{\mathrm{cc}}}
\newcommand{\bcc}{\mathop{\mathrm{bcc}}}
\newcommand{\fneg}{\mathop{\mathrm{neg}}}
\newcommand{\tr}{\mathop{\mathrm{tr}}}

\newcommand{\CONT}{\textup{\textsf{CONT}}}
\newcommand{\STABLE}{\textup{\textsf{STABLE}}}
\newcommand{\UNSTABLE}{\textup{\textsf{UNSTABLE}}}
\newcommand{\MONO}{\textup{\textsf{MONO}}}
\newcommand{\SDOM}{\textup{\textsf{SDOM}}}

\begin{document}

\title{On the Expressive Power of First-Order Boolean Functions in
PCF\thanks{This paper is essentially the same as one that appeared in
\emph{Theoretical Computer Science} 266(1-2), pp. 543-567, 2001.
This work was done while the first author was at McGill
University, and was supported in part by a scholarship from
FCAR. A preliminary version of this paper was written while the first
author was at Bell Laboratories, Lucent Technologies.}}
\author{Riccardo Pucella\\
Department of Computer Science\\
Cornell University\\
Ithaca, NY, 14853, USA
\and
Prakash Panangaden\\
School of Computer Science\\
McGill University\\
Montreal H3A 2K7, Canada
}
\date{}

\maketitle

\begin{abstract}
Recent results of Bucciarelli show that the semilattice of degrees of
parallelism of first-order boolean functions in PCF has both infinite
chains and infinite antichains. By considering a simple subclass
of Sieber's sequentiality relations, we identify levels in the
semilattice and derive inexpressibility results concerning functions on
different levels. This allows us to further explore the structure of
the semilattice of degrees of parallelism: we identify semilattices
characterized by simple level properties, and show the existence of
new infinite hierarchies which are in a certain sense natural with
respect to the levels.
\end{abstract}

\section{Introduction}

In this paper we study the relative definability of  first-order
boolean functions with respect to Plotkin's language PCF \cite{Plotkin77},
a simply-typed $\lambda$-calculus with recursion over the ground
types of integers and booleans. Relative definability defines a preorder on continuous boolean
functions, and this ordering induces a natural equivalence relation. The object of
our study will be the structure of the resulting partially ordered set of
equivalence classes of functions (called degrees of parallelism). Work by
Trakhtenbrot \cite{Trakhtenbrot73,Trakhtenbrot75}, Sazonov
\cite{Sazonov76}, Lichtenth\"{a}ler \cite{Lichtenthaler96} and Bucciarelli and Malacaria
\cite{Bucciarelli97,UNSTABLE:Bucciarelli97} 
show that the structure of degrees of parallelism is highly
non-trivial: even when restricted to first-order functions, the poset
forms a sup-semilattice and contains a "two-dimensional" hierarchy of
functions, both infinite chains and infinite antichains of functions.

It is known that the definability ordering is completely
characterized by the sequentiality relations of Sieber. The result is
a duality of sorts: $f$ can be defined using $g$ if the sequentiality
relations under which $g$ is invariant is a subset of the
sequentiality relations under which $f$ is invariant. Therefore, it
seems worthwhile to try to derive the set of sequentiality relations
under which a given function is invariant. As a first step towards
this goal we focus our attention in this paper on a simple
class of sequentiality relations we call presequentiality
relations. Invariance under presequentiality relations induces a
coarser ordering on functions than full sequentiality relations, from
which we cannot infer definability results but can infer strong
inexpressibility results. In effect, this coarser ordering is a
``skeleton'' of the definability preorder. 

The main advantage of working with presequentiality relations is that
we can completely characterize the set of presequentiality relations
under which a given function is invariant. It turns out that a pair of 
integers is sufficient to completely describe this set. This pair of
integers, called the presequentiality level of the function, can 
straightforwardly be derived from the trace of the function. Well-known
functions in the definability preorder, such as Parallel OR, the
Berry-Plotkin function, the Gustave function, the Detector function, can be easily
characterized in terms of presequentiality levels. We use
presequentiality levels to guide our exploration of the 
definability preorder: we present subsemilattices with natural
presequentiality level characterizations, namely the stable, unstable, 
stable-dominating and monovalued functions. We exhibit
natural hierarchies of functions in these lattices, where natural
is taken to mean that every function in the hierarchy has a different
presequentiality level,
thereby making the hierarchy part of the skeleton of the definability
preorder. 

This paper is structured as follows. In the next section, we review
the required mathematical preliminaries, rigorously defining the
notions of relative definability, traces, linear coherence, as well as 
stating useful existing results. In Section
\ref{s:preseq}, we study presequentiality relations, and prove the two 
main lemmas of this paper: the Reduction Lemma and the Closure Lemma,
which allow us to find canonical representatives for the set of
presequentiality relations under which a function is invariant. In
Section \ref{s:plevels}, we point out the relationship between the
canonical representatives and the trace of the function, and thus 
define the notion of presequentiality level. Section \ref{s:struct} then investigates 
the structure of the definability preorder guided by
presequentiality levels, as described above. 

This work is in the lineage of the work of Bucciarelli in 
\cite{Bucciarelli97} and Bucciarelli and Malacaria in
\cite{UNSTABLE:Bucciarelli97}. The main results from this paper were
originally reported in \cite{Pucella96}.

\section{Preliminaries}
\label{s:prelim}

In this section, we review some of the mathematical background to our study of
first-order monotone boolean functions and the previous work 
already done on the subject by Trakhtenbrot, Sazanov, Bucciarelli and
Malacaria. We assume knowledge of PCF and its continuous model
\cite{Plotkin77}, as well as a passing familiarity with logical
relations \cite{Plotkin80}. Let \bool\ be the flat
domain of boolean values. Given \boolfn{f}{k}\ and $x =
\indexedtuple{x}{k}$, then $f(x)$ 
stands 
for \appliedsub{f}{x}{k}, and given $A= \indexedsupset{x}{n}
\subseteq \boolm{k}$, $f(A)$ is defined to be $\left\{
f(x^i) : x^i \in A\right\}$. As usual, $\pi_{1}$ and $\pi_{2}$
represent the projection functions associated with the cartesian
product on sets.  

Relative definability refers to the ability to define some function 
using another function: a function can define another function if
there exist some algorithm in some language that uses the former to
compute the latter. In our case, algorithms are taken to be PCF-terms: 
given two continuous functions $f$ and $g$, we say that $f$ is
\emph{PCF-expressible} (or simply expressible) by $g$, denoted $f \parleq\ g$, if there exists a PCF-term $M$
such that $f = \Mean{M}g$. Equivalent terminologies in the literature for $f\parleq g$ 
are ``f is less parallel than g'', or $f$ is $g$-expressible. The
\parleq\ preorder induces an equivalence 
relation \pareq\ on continuous function such that $f\pareq g$ iff $f
\parleq g$ and $g \parleq f$. The equivalence classes are 
called \emph{degrees of parallelism}, and two functions $f$,$g$
with $f \pareq g$ are called \emph{equiparallel}. The degree of
parallelism of a continuous function $f$ is denoted [$f$]. 

We are interested in studying the structure of first-order degrees of
parallelism. Trakhtenbrot \cite{Trakhtenbrot73,Trakhtenbrot75} and Sazonov
\cite{Sazonov76} first investigated the subject and pointed out finite
subposets of degrees (though not necessarily first-order
degrees). Some facts are consequences of well-known
results. The poset of degrees of parallelism must have a top element,
Parallel OR (POR), by Plotkin's full abstraction result for PCF+POR
\cite{Plotkin77}. On the other hand, the poset must have a bottom
element, the degree of all M-sequential functions. Indeed, a
fundamental property of PCF is that PCF-definable 
functions are exactly the M-sequential functions. A function
\boolfn{f}{k} is \emph{M-sequential} \cite{Milner77} (or simply
sequential) if it is
constant or if there exists an  integer $i$ (called an \emph{index of
sequentiality}) with $1 \leq i \leq k$ such that $x_{i} = \bot$
implies that $f(x_{1},\ldots,x_{k}) = \bot$ and such that for any
fixed value $x_{i}$, the function of the remaining arguments is also
M-sequential. In \cite{UNSTABLE:Bucciarelli97}, it is proved that
first-order degrees of parallelism form a sup-semilattice, which we
will denote \CONT\footnote{
  \CONT\ refers to the fact that those functions are continuous:
  recall that for first-order boolean functions, monotone functions
  are continuous.
}.
\begin{prop}
The poset of first-order degrees of parallelism is a sup-semilattice
with a bottom element (the set of sequential functions) and a top
element (the degree of \funpor).
\end{prop}

The trace of a function is the central notion we use to study boolean
functions. The trace is a representation of the minimum inputs needed for the
function to produce a result. Formally, given a first-order monotone function
\boolfn{f}{k}, the \emph{trace} of $f$ is \[\tr(f) = \left\{(v,b) |
v\in \boolm{k}, b\in \bool, b\not= \bot, f(v)=b \mbox{ and } \forall
v'<v, f(v')=\bot \right\} \] For $x,y\in\bool$, let $x\uparrow y$ hold 
if
$x$ and $y$ have a common upperbound in $\bool$, that is if $x=\bot$ or
$y=\bot$ or $x=y$. Extend $\uparrow$ pointwise to tuples in
$\boolm{n}$. It is easy to see that a first-order monotone boolean
function $f$ is stable (in the sense of Berry \cite{Berry76}) if and
only if for all $v_{1},v_{2} \in \pi_{1}(\tr(f))$, $v_{1} \not\uparrow
v_{2}$. Note that the monotonicity of $f$ insures that if $v_{1}
\uparrow v_{2}$ then $f(v_{1}) = f(v_{2})$. For a set of tuples
$A\subseteq\boolm{k}$, a set $B\subseteq\boolm{k}$ is an Egli-Milner
lowerbound for $A$ if for every $x\in A$, there is a $y\in B$ with
$y\leq x$, and for every $y\in B$, there is an $x\in A$ with $y\leq x$.

Linear coherence is used by Bucciarelli and Erhard to study
first-order boolean functions in
\cite{Bucciarelli91,Bucciarelli94,Bucciarelli97}. A subset
$A=\indexedsupset{v}{n}$ of $\boolm{k}$ is \emph{linearly coherent} (or
simply coherent) if for every coordinate, either a tuple in $A$
contains $\bot$ at that coordinate, or all the tuples in $A$ have
the same value at that coordinate, that is
\[ \forall j\in\{1,\ldots,k\} \left(\forall l\in\{1,\ldots,n\},v^{l}_{j}
\not= \bot \right) \Rightarrow \forall l_{1},l_{2}\in\{1,\ldots,n\}, v_{j}^{l_{1}}=v_{j}^{l_{2}} \] A subset
$A=\indexedsupset{v}{n}$ of \boolm{k} is \emph{\bottom-covering} if for
every coordinate a tuple in $A$ contains $\bot$ at that
coordinate, that is
\[ \forall j\in\{1,\ldots,k\}, \exists i\in\{1,\ldots,k\}, v_{j}^{i} = \bot \]
It is easy to see that if $A$ is \bottom-covering then $A$ is
coherent. Abusing the terminology, we will sometimes say that a
first-order monotone boolean
function $f$ is \bottom-covering if $\pi_{1}(\tr(f))$ has the
corresponding property. 

Monovalued functions are an important class of functions we study. 
A first-order monotone boolean function $f$ is \emph{monovalued} if
$\setsize{\pi_{2}(\tr(f))}=1$. 
By another abuse of terminology, we will say that a subset $A
\subseteq \pi_{1}(\tr(f))$ is monovalued if $\setsize{f(A)} = 1$. A
boolean function which is not monovalued will sometimes be called
bivalued\footnote{
  The term ``bivalued'' refers of course to the
  fact that there are two non-$\bot$ values in the boolean domain ---
  a function is bivalued if $\setsize{\pi_2(\tr(f))}=2$.
}.

We define two operations on boolean functions. Given a
first-order monotone boolean function \boolfn{f}{k}, let \boolfn{\fneg(f)}{k} be the
function returning \true\ when $f$ returns \false\ and returning \false\
when $f$ returns \true. As for 
the second operation, given two first-order monotone boolean functions
\boolfn{f}{k} and \boolfn{g}{k'}, 
(without loss of generality, assume there exists an $l\geq 0$ with
$k=k'+l$) define the function \boolfn{f+g}{\max(k,k')+1} by the
following trace: 
\begin{eqnarray*}
\tr(f+g) & = & \{((\true,x_{1},\ldots,x_{k}),b) :
((x_{1},\ldots,x_{k}),b) \in \tr(f)\} \bigcup \\
& &
\{((\underbrace{\false,\ldots,\false}_{l+1},x_{1},\ldots,x_{k'}),b)
: ((x_{1},\ldots,x_{k'}),b) \in \tr(g)\}
\end{eqnarray*}
As shown in \cite{UNSTABLE:Bucciarelli97}, $f+g$ is equiparallel to the least upperbound of
$f$ and $g$ in \CONT, in other words $[f+g] = [f] \vee [g]$. 

Bucciarelli illustrates the non-trivial structure of the \CONT{} semilattice
by exhibiting hierarchies\footnote{
  A hierarchy is simply an $\omega$-chain in the definability preorder.
} of functions in \CONT{}
\cite{Bucciarelli97}. He defines the function \bucc{n}{m} via
the following description: the trace of \bucc{n}{m} has $m$ elements and
each trace element returns \true; for any subset of less than $n$ elements
(and at least two) of the first projection of the trace, there exists a
coordinate which makes that subset incoherent. The Bucciarelli hierarchy
is actually a two-dimensional infinite hierarchy of
functions. 

Generalizing the techniques used in \cite{Bucciarelli97}, Bucciarelli
and Malacaria prove the following proposition in
\cite{UNSTABLE:Bucciarelli97}, in their attempt to find a
characterization of the \CONT{} semilattice in terms of hypergraphs
(this proposition is restated so that it does not refer to hypergraphs)
\begin{prop}[Bucciarelli, Malacaria]
\label{p:buccmala}
Let $f,g$ be two first-order monotone boolean functions. If there exists a function
\function{\alpha}{\tr(f)}{\tr(g)} such that
\begin{enumerate}
\item for all $A \subseteq \tr(f)$, if $\pi_{1}(A)$ is non-singleton and
linearly coherent, then $\pi_{1}(\alpha(A))$ is non-singleton and
linearly coherent.
\item for all $A \subseteq \tr(f)$ with $\pi_{1}(A)$ non-singleton
and linearly coherent, and for all $x,y \in A$, we have $\pi_{2}(x)
\not= \pi_{2}(y) \Rightarrow \pi_{2}(\alpha(x)) \not=
\pi_{2}(\alpha(y))$.
\end{enumerate}
then $f \parleq g$.
\end{prop}

This property will be used often in this paper to prove definability
results between functions.

\section{Presequentiality relations}
\label{s:preseq}

Relative definability for first-order boolean functions is fully
characterized by Sieber's sequentiality relations, introduced in
\cite{Sieber92}. Sequentiality relations are the logical relations
\cite{Plotkin80} under which the constants of PCF are
invariant. Recall that an $n$-ary logical relation $R$ on a
$\lambda$-model $(D^\tau)_{t\in\mbox{\scriptsize Type}}$ is a family of relations
$R^\tau\subseteq(D^\tau)^n$ such that for all types $\sigma,\tau$ and
$f_1,\ldots,f_n\in D^{\sigma\rightarrow\tau}$,
\[R^{\sigma\rightarrow\tau}(f_1,\ldots,f_n)\Leftrightarrow\forall
d_1,\ldots,d_n,R^\sigma(d_1,\ldots,d_n)\Rightarrow R^\tau(f_1
d_1,\ldots, f_n d_n)\] An element $d\in D^\tau$ is \emph{invariant}
under $R$ if $R^\tau(d,\ldots,d)$ holds. We now give the definition of
sequentiality relations in a slightly different form than Sieber in
\cite{Sieber92}, distinguishing the simple kind of sequentiality
relations which we call presequentiality relations. 
\begin{defn}
For each $n\geq 0$ and each pair of sets $A \subseteq B \subseteq
\left\{1,\ldots,n\right\}$, the \emph{presequentiality relation}
$\preseq{A}{B}{n}\subseteq (D^{\tau})^{n}$, $\tau=\iota,o$, is
an $n$-ary logical relation defined by 
\[ \preseq{A}{B}{n}\left(d_{1},\ldots,d_{n}\right) \miff (\exists i\in
A. d_{i} = \bot) \vee (\forall i,j \in B. d_{i} = d_{j}) \]
An $n$-ary logical relation $R$ is called a \emph{sequentiality relation} if
$R$ is an intersection of presequentiality relations.
\end{defn}

Sieber's relations are defined for full PCF, that is with both
integers (type $\iota$) and booleans (type $o$). For the purposes of
this paper, it is sufficient to look at relations over the booleans,
that is over $\bool = D^o$. For the special case of a first-order
boolean function $\boolfn{f}{k}$, invariance under \preseq{A}{B}{n}
means that for tuples
$(x^1_1,\ldots,x^1_n),\ldots,(x^k_1,\ldots,x^k_n)$ in
\preseq{A}{B}{n}, we have \appliedtuple{f}{x}{k}{n} also in \preseq{A}{B}{n}.
The following proposition, proved in \cite{Sieber92}, gives the full
characterization of the definability preorder for first-order functions. It is interesting to note that this characterization is effective and
Stoughton implemented an algorithm that decides $f
\parleq g$ given the functions $f$ and $g$ \cite{Stoughton94}.
\begin{prop}[Sieber]
\label{p:sieber}
For any first-order monotone boolean functions $f$ and $g$, $f\parleq\
g$ if and only if for any sequentiality relation $R$, if $g$ 
is invariant under $R$ then $f$ is also invariant under $R$.
\end{prop}

Proposition \ref{p:sieber} tells us that a function $f$ is not
$g$-expressible if we can exhibit a sequentiality relation $R$ such
that $g$ is invariant under $R$ but $f$ is not. If we restrict our
attention to presequentiality relations, it is easy to see that
invariance under presequentiality relations induces a coarser ordering
than invariance under sequentiality relations, that is it identifies more functions. If two functions are invariant under the
same presequentiality relations, then nothing can be said about their
relative definability. However, if they are not invariant under the
same presequentiality relations, we can derive strong inexpressibility 
results, since presequentiality relations are a weak class of
sequentiality relations. In effect, invariance under presequentiality
relations can be viewed as defining the ``skeleton'' of the relative
definability preorder. The advantage of working with presequentiality
relations is that they are simpler than full sequentiality relations,
and a great deal of structure can be extracted straightforwardly, as
we will presently see.

The central problem of
this paper is to determine the presequentiality
relations under which a given function is invariant. An early
restricted form of this may already be found in
\cite{Bucciarelli97}. The following two lemmas show that it is not
necessary to consider every 
presequentiality relation. The Reduction Lemma tells us that it is
sufficient to look at presequentiality relations of a simple form.
The Closure Lemma says that if a function is invariant under a
presequentiality relation \preseq{A}{B}{n}, invariance holds under any
presequentiality relation with ``smaller'' $A$ and $B$. In Section \ref{s:plevels}, we will see how these lemmas lead to a simple
characterization of the set of presequentiality relations under which
a function is invariant. 

\begin{lem}[Reduction Lemma]
Given \boolfn{f}{k} a first-order monotone boolean function and $A \subseteq B \subseteq
\seqsetto{n}$, one of the following holds:
\begin{enumerate}
\item $(A=B)$ $f$ is invariant under \preseq{A}{A}{n} \miff
$f$ is invariant under
\preseqseq{\setsize{A}}{\setsize{A}}{\setsize{A}}
\item $(A\subset B)$ $f$ is invariant under \preseq{A}{B}{n}
\miff $f$ is invariant under
\preseqseq{\setsize{A}}{\setsize{A}+1}{\setsize{A}+1}.
\end{enumerate}
\end{lem}

\begin{lem}[Closure Lemma]
Given \boolfn{f}{k} a first-order monotone boolean function and $m$ any integer with $m
\geq 0$, the following holds:
\begin{enumerate}
\item $f$ invariant under \preseqseq{m}{m+1}{m+1} $\Rightarrow$ $f$
invariant under \preseqseq{m}{m}{m}.
\item $f$ invariant under \preseqseq{m+1}{m+1}{m+1}
$\Rightarrow$ $f$ invariant under \preseqseq{m}{m}{m}
\item $f$ invariant under \preseqseq{m+1}{m+2}{m+2}
$\Rightarrow$ $f$ invariant under \preseqseq{m}{m+1}{m+1}
\end{enumerate}
\end{lem}

The proof of these lemmas is much more digestible when split across
several technical lemmas (\ref{l:3},\ref{l:5},\ref{l:8}) which we now state and prove. 

\begin{lem}
\label{l:3}
Let $m(M)$ be the least $n$ such that $M\subseteq \seqsetto{n}$, and
let \boolfn{f}{k} be a first-order monotone boolean function. The function  $f$ is
invariant under \preseq{A}{B}{n} iff $f$ is invariant under
\preseq{A}{B}{m(B)}. 
\end{lem}
\begin{pf} 
$(\Rightarrow)$ We show that if $f$ is invariant under
\preseq{A}{B}{n}, then for all $n'\leq n$ such that 
$B \subseteq \seqsetto{n'}$, $f$ is invariant under
\preseq{A}{B}{n'}. 

\begin{sloppypar}
For the sake of contradiction, assume there exist $n, A, B,
n'$ with $n'\leq n$ such that $f$ is invariant under
\preseq{A}{B}{n} but not under 
\preseq{A}{B}{n'}. That is, there exist tuples 
$\indexedtuple{{x}^{1}}{n'}, \cdots, \indexedtuple{{x}^{k}}{n'} \in
\preseq{A}{B}{n'}$
and \tuplenotin{\indexedtuple{y}{n'}}{\preseq{A}{B}{n'}}
with $y_{i} = \applied{f}{x_{i}}{k}$.
\end{sloppypar}

\begin{sloppypar}
The tuples
\[(x^{1}_{1},\ldots,x^{1}_{n'},\bottom,\ldots,\bottom),\cdots,
 (x^{k}_{1},\ldots,x^{k}_{n'},\bottom,\ldots,\bottom) \] 
then must be in \preseq{A}{B}{n}. Since
\tuplenotin{(y_1,\ldots,y_{n'})}{\preseq{A}{B}{n'}}, we must have \tuplenotin{(y_{1},\ldots,y_{n'},\bottom,\ldots,\bottom)}
               {\preseq{A}{B}{n}}, 
contradicting the invariance of $f$ under
\preseq{A}{B}{n}. 
\end{sloppypar}

($\Leftarrow$) We show that if $f$ is invariant under
\preseq{A}{B}{n}, then for all $n'\geq n$,  $f$ is 
invariant under \preseq{A}{B}{n'}.

\begin{sloppypar}
For the sake of contradiction, assume there exist $n, A, B$ and $n'
\geq n$ such that $f$ is invariant under \preseq{A}{B}{n}
but not under \preseq{A}{B}{n'}. That is, there exist tuples
\tuplearray{x}{n'}{k}{\preseq{A}{B}{n'}}
and
\tuplenotin{\indexedtuple{y}{n'}}{\preseq{A}{B}{n'}}
with $y_{i} = \applied{f}{x_{i}}{k}$.
Observe that
$\tuplein{\indexedtuple{x}{n'}}{\preseq{A}{B}{n'}}
\miff
\tuplein{\indexedtuple{x}{n}}{\preseq{A}{B}{n}}$. 
Hence, 
\tuplearray{x}{n}{k}{\preseq{A}{B}{n}}
but
\tuplenotin{\indexedtuple{y}{n}}{\preseq{A}{B}{n}}
contradicting the invariance of $f$ under \preseq{A}{B}{n}. \qed
\end{sloppypar}
\end{pf}

\begin{lem}
\label{l:5}
Given \boolfn{f}{k} a first-order monotone boolean function, $f$ is invariant under
\preseq{A}{B}{n} iff $f$ is invariant under 
\preseqseq{\setsize{A}}{\setsize{B}}{n}.
\end{lem}
\begin{pf} 
We show the following more general result: let $A,B,C,D$ be sets
with $A\subseteq B\subseteq \seqsetto{n}, C\subseteq D\subseteq
\seqsetto{n}$, and let $p$ be a permutation of $\seqsetto{n}$ into
$\seqsetto{n}$ such that $p(A) = C$ and $p(B) = D$. Then $f$ is
invariant under \preseq{A}{B}{n} \miff $f$ is invariant under
\preseq{C}{D}{n}. 

Let us first prove that 
\begin{equation}
 \tuplein{\indexedtuple{x}{n}}{\preseq{A}{B}{n}} \miff
\tuplein{(x_{p^{-1}(1)},\ldots,x_{p^{-1}(n)})}{\preseq{C}{D}{n}}.
\label{e:simplify}
\end{equation}
Let \tuplein{\indexedtuple{x}{n}}{\preseq{A}{B}{n}}, and
$y_{i} = x_{p^{-1}(i)}$. To show
\tuplein{\indexedtuple{y}{n}}{\preseq{C}{D}{n}}, consider the two cases:
\begin{enumerate}
\item There is an $i\in A, x_{i} = \bottom$. In which case, let $c =
  p(i)$, with $c\in C$ since $i\in A$. Moreover, $y_{c} = x_{p^{-1}(c)}
  = x_{p^{-1}(p(i))} = x_{i} = \bottom$, so there is a $j \in C,
  y_{j}=\bottom$.
\item For all $i,j\in B, x_{i} = x_{j}$. Assume there are $i,j \in D,
y_{i} \not = y_{j}$. Then $x_{p^{-1}(i)} \not = x_{p^{-1}(j)}$, hence
there are $i',j' \in B, x_{i'} \not =
x_{j'}$, a contradiction. Hence for all $i,j \in D, y_{i} =
y_{j}$.
\end{enumerate}
Hence \tuplein{\indexedtuple{y}{n}}{\preseq{C}{D}{n}}. The
reverse direction follows by symmetry of the permutation $p$, proving (\ref{e:simplify}).

Now, observe that we need only show one
direction of the general result (the 
reverse direction follows by symmetry of the permutation $p$).

Consider any tuples
\tuplearray{x}{n}{k}{\preseq{A}{B}{n}}. 
Let $y_{i} = \applied{f}{x_{i}}{k}$. Since $f$ is invariant under
\preseq{A}{B}{n}, \tuplein{\indexedtuple{y}{n}}{\preseq{A}{B}{n}}. 

By (\ref{e:simplify}),  each tuple \indexedtuple{x^{j}}{n} is also in
\preseq{C}{D}{n} and so is
\tuplein{\indexedtuple{y}{n}}{\preseq{C}{D}{n}}, hence $f$ is
invariant under \preseq{C}{D}{n}. 

To prove the lemma, it is sufficient to show that there exists a
permutation $p$ of \seqsetto{n} such that
$p(A)=\seqsetto{\setsize{A}}$, $p(B)=\seqsetto{\setsize{B}}$, which is 
immediate. \qed
\end{pf}

\begin{lem}
\label{l:8}
Given \boolfn{f}{k} a first-order monotone boolean function. Then $f$ is invariant under
\preseq{A}{B}{n}, $\setsize{B \backslash A}=1$ iff $f$ is 
invariant under \preseq{A}{B'}{n} for any
$B'$ such that $B \subseteq B'$.
\end{lem}
\begin{pf}
($\Rightarrow$) We show that if $f$ is invariant under
\preseq{A}{B}{n}, $\setsize{B \backslash A}=1$, then for 
any $B'$ such that $B \subseteq B'$, $f$ is
invariant under \preseq{A}{B'}{n}.

By Lemma \ref{l:3} and Lemma \ref{l:5}, it is sufficient to show that
for any $m$, if $f$ invariant under \preseqseq{m}{m+1}{m+1} then $f$
is invariant under \preseqseq{m}{n}{n} for any $n\geq m+1$. 

For the sake of contradiction, assume that for some $m$ and $n\geq
m+1$, $f$ is invariant under the presequentiality relation
\preseqseq{m}{m+1}{m+1} but not under
\preseqseq{m}{n}{n}. Then there are tuples
\tuplearray{x}{n}{k}{\preseqseq{m}{n}{n}}
but \tuplenotin{\indexedtuple{y}{n}}{\preseqseq{m}{n}{n}}, for
$y_{i} = \applied{f}{x_{i}}{k}$.
Hence, for all $i\leq m$, $y_{i} \not= \bot$ and there are $I,J$ such
that $y_{I} \not= y_{J}$. Without loss of generality, choose $I$ the
minimal such index. 

We proceed by case analysis on the value of $I$ and $J$: 
\begin{enumerate}
\item $(I \leq m)$ Consider the following tuples
$\left(x_{1}^{1},\ldots,x_{m}^{1},x_{J}^{1}\right),\cdots,\left(x_{1}^{k},\ldots,x_{m}^{k},x_{J}^{k}\right)$ 
which are in \preseqseq{m}{m+1}{m+1}; by assumption of the
invariance of $f$, we have
\tuplein{\left(y_{1},\ldots,y_{m},y_{J}\right)}{\preseqseq{m}{m+1}{m+1}}
Hence, either there is $i\leq m$ such $y_{i} =\bot$ (a contradiction), 
or $y_{I}=y_{J}$ (also a contradiction). 

\item $(J \leq m)$ Same argument.

\item $(I,J > m)$ We further consider 3 subcases.
\begin{enumerate}
\item 
($y_{I}=\bot$). Consider the %
tuples
$\left(x_{1}^{1},\ldots,x_{m}^{1},x_{I}^{1}\right),\dots,\left(x_{1}^{k},\ldots,x_{m}^{k},x_{I}^{k}\right)$ 
which are in \preseqseq{m}{m+1}{m+1}; by assumption of the
invariance of $f$ we have
\tuplein{\left(y_{1},\ldots,y_{m},y_{I}\right)}{\preseqseq{m}{m+1}{m+1}}.
So either there is $i \leq m$ such that $y_{i} = \bot$ (a
contradiction), or $y_{I}=y_{i}$ for all $i \leq m$ (also a
contradiction) 

\item ($y_{J}=\bot$) Same argument.

\item ($y_{I},y_{J} \not= \bot$) By choice of minimal $I$, we know
that $y_1=\cdots=y_m$ and all are either \true\ or \false. On the other 
hand, $y_I\not= y_J$ and $y_I,y_J\not=\bot$, so let $c=$ $I$ or $J$,
such that $y_c\not= y_1$. 

Consider the tuples
$\left( x_{1}^{1},\ldots,x_{m}^{1},x_c^{1}\right),\cdots,\left(
x_{1}^{k},\ldots,x_{m}^{k},x_c^{k}\right)$, 
easily seen to be tuples 
in
\preseqseq{m}{m+1}{m+1}, and by assumption of the invariance of $f$,
we have 
\tuplein{\left(y_{1},\ldots,y_{m},y_c\right)}{\preseqseq{m}{m+1}{m+1}}.
So either there is an $i \leq m$ such that $y_{i} = \bot$ (a
contradiction), or $y_c=y_1$ (also a contradiction)
\end{enumerate}
\end{enumerate}

$(\Leftarrow)$ We show that if $f$ is invariant under
\preseq{A}{B}{n}, then $f$ is invariant under 
\preseq{A}{B'}{n} for all $A\subseteq B' \subseteq B$.

For the sake of contradiction, assume there exist $n, A, B, B'$ with
$A\subseteq B' \subseteq B$ such that $f$ is invariant under
\preseq{A}{B}{n} but not under
$\preseq{A}{B'}{n}$. Then there exist tuples
\tuplearray{x}{n}{k}{\preseq{A}{B'}{n}}
such that \tuplenotin{\indexedtuple{y}{n}}{\preseq{A}{B'}{n}}
with $y_{i} = \applied{f}{x_{i}}{k}$.

Fix an arbitrary $I \in A$. Consider the following tuples:
\indexedtuple{z^{j}}{n} for $1 \leq j \leq k$, with
\[ z^{j}_{i} = \left\{ \begin{array}{ll}
               x^{j}_{i} & \mbox{if $i \in B'$} \\
               x^{j}_{I} & \mbox{if $i \in B\backslash B'$} \\
               \bottom & \mbox{otherwise}
	               \end{array}
               \right. \]

We first verify that these tuples are in \preseq{A}{B}{n}. For each $j, 1
\leq j \leq k$, consider the original tuple
\tuplein{\indexedtuple{x^{j}}{n}}{\preseq{A}{B'}{n}}. In other
words, either
\begin{enumerate}
\item there is an $i\in A$, $x^{j}_{i} = \bottom$, and for that
  $i\in A$, we have $z^{j}_{i} = x^{j}_{i} = \bottom$. Hence
  \tuplein{\indexedtuple{z^{j}}{n}}{\preseq{A}{B}{n}}, or
\item For all $i\in A$, $x^{j}_{i} \not= \bottom$, and for all $
  i,i'\in B', x^{j}_{i} = x^{j}_{i'}$. Hence, for all $
  i,i'\in B', z^{j}_{i} = z^{j}_{i'}$. Moreover, for all $i\in
  B\backslash B', z^{j}_{i} = x^{j}_{I}$ for $I \in A \subseteq
  B'$. Hence, for all $i,i' \in B, z^{j}_{i} =
  z^{j}_{i'}$ and the tuple 
\tuplein{\indexedtuple{z^{j}}{n}}{\preseq{A}{B}{n}}.
\end{enumerate}

By the above construction, we see that for all $i \in B',
\applied{f}{z_{i}}{k} = y_{i}$ .

Since \tuplenotin{\indexedtuple{y}{n}}{\preseq{A}{B'}{n}}, we
have for all $i\in A, y_{i} \not= \bottom$ and there are $i,j\in
B', y_{i} \not=y_{j}$. This implies that for all $i\in A,
\applied{f}{z_{i}}{k} \not= \bottom$ and there are $i,j \in
B'\subseteq B$ such that $\applied{f}{z_{i}}{k} \not=
\applied{f}{z_{j}}{k}$. In other words, $f$ is not invariant under
\preseq{A}{B}{n}, contracting the assumption.\qed

\end{pf}

The proofs of the Reduction and Closure Lemmas are now immediate.
\vspace{0.2in}

\begin{pf}(Reduction Lemma)
\begin{enumerate}
\item $(A=B)$ By Lemma \ref{l:5}, we have that $f$ is invariant
under \preseq{A}{A}{n} iff $f$ is invariant under
\preseqseq{\setsize{A}}{\setsize{A}}{n} and by Lemma 
\ref{l:3}, $f$ is invariant under
\preseqseq{\setsize{A}}{\setsize{A}}{n} iff $f$ is invariant under
\preseqseq{\setsize{A}}{\setsize{A}}{\setsize{A}}. 

\item \begin{sloppypar}
$(A \subset B)$ By Lemma \ref{l:5}, $f$ is
invariant under \preseq{A}{B}{n} iff $f$ is invariant under
\preseqseq{\setsize{A}}{\setsize{B}}{n}. By Lemma 
\ref{l:8}, $f$ is invariant under
\preseqseq{\setsize{A}}{\setsize{B}}{n} iff $f$ is invariant under
\preseqseq{\setsize{A}}{\setsize{A}+1}{n}, and by Lemma 
\ref{l:3}, this happens iff $f$ is invariant under
\preseqseq{\setsize{A}}{\setsize{A}+1}{\setsize{A}+1}. \qed
      \end{sloppypar}
\end{enumerate} 
\end{pf}

\begin{pf}(Closure Lemma)
\begin{enumerate}
\item The ($\Leftarrow$) direction in the proof of Lemma \ref{l:8}
actually proves this case.

\item Given tuples
\tuplearray{x}{m}{k}{\preseqseq{m}{m}{m}}
we show \tuplein{\indexedtuple{y}{m}}{\preseqseq{m}{m}{m}} with
$y_{i} = \applied{f}{x_{i}}{k}$.

By assumption, the tuples
$\left(x_{1}^{1},\ldots,x_{m}^{1},x_{1}^{1}\right),\cdots,\left(x_{1}^{k},\ldots,x_{m}^{k},x_{1}^{k}\right)$ 
are in $\preseqseq{m+1}{m+1}{m+1}$. 

\begin{sloppypar}
By invariance of $f$ under \preseqseq{m+1}{m+1}{m+1}, we
have $\left(y_{1},\ldots,y_{m},y_{1}\right) \in
\preseqseq{m+1}{m+1}{m+1}$ which means that either there is $i\leq m $
such that $y_{i} = \bot$ or for all $i,j \leq m$, $y_{i} =
y_{j}$. Hence $\tuplein{\indexedtuple{y}{m}}{\preseqseq{m}{m}{m}}$.
\end{sloppypar}

\item Same argument as part (2): assume tuples
$\left(x_1^i,\ldots,x_{m+1}^i\right)$ in \preseqseq{m}{m+1}{m+1}, and
consider the tuples $\left(x_1^i,\ldots,x_m^i,x_1^i,x_{m+1}^i\right)$. \qed
\end{enumerate}

\COMMENTOUT{Given the tuples 
\tuplearray{x}{m+1}{k}{\preseqseq{m}{m+1}{m+1}}
we show \tuplein{\indexedtuple{y}{m+1}}{\preseqseq{m}{m+1}{m+1}}, with
$y_{i} = \applied{f}{x_{i}}{k}$.

Consider the tuples
$\left(x_{1}^{1},\ldots,x_{m}^{1},x_{1}^{1},x_{m+1}^{1}\right),\cdots,\left(x_{1}^{k},\ldots,x_{m}^{k},x_{1}^{k},x_{m+1}^{k}\right) \in  \preseqseq{m+1}{m+2}{m+2}$. 

By invariance of $f$ under \preseqseq{m+1}{m+2}{m+2}, we have
\[\left(y_{1},\ldots,y_{m},y_{1},y_{m+1}\right) \in
\preseqseq{m+1}{m+2}{m+2}\] which means either that there is $i\leq m$
such that $y_{i}=\bot$ or for all $i,j\leq m+1$, $y_{i}=y_{j}$. Hence,
\[\tuplein{\indexedtuple{y}{m+1}}{\preseqseq{m}{m+1}{m+1}}. \qed \] }

\end{pf}

\section{Presequentiality levels}
\label{s:plevels}

The Reduction Lemma and the Closure Lemma of the previous section can be used
to show that the set of presequentiality relations under which a
function is invariant is characterized by two integers (allowing
for $\infty$). Given $f$ a function invariant under presequentiality
relations $\{\preseq{A_i}{B_i}{n}\}_{i\in I}$; by the Reduction Lemma, 
this is equivalent to saying that $f$ is invariant under the
presequentiality relations
$\{\preseqseq{\setsize{A_i}}{\setsize{A_i}}{\setsize{A_i}}\}_{i\in I,
A_i = B_i}$ and $\{\preseqseq{\setsize{A_i}}{\setsize{A_i}+1}{\setsize{A_i}+1}\}_{i\in
I,A_i\subset B_i}$. By the Closure Lemma, there must exist maximal $i$ 
and $j$ (possibly $\infty$) such that $f$ is invariant under
$\preseqseq{k}{k}{k}$ for all $k\leq i$ and $f$ is invariant under
$\preseqseq{k}{k+1}{k+1}$ for all $k\leq j$. We will call the pair
$(i,j)$ the \emph{presequentiality level} (p-level) of the function
$f$. Clearly, a function with a p-level of $(\infty,\infty)$ is
invariant under all presequentiality relations. Since every function
in a degree of parallelism must be invariant under the same presequentiality
relations (by Proposition \ref{p:sieber}), we also talk about
the presequentiality level of a degree of parallelism. Alternatively,
a function with a p-level of $(i,j)$ is easily seen by applications of 
the Reduction Lemma and the Closure Lemma to be invariant under a
presequentiality relation \preseq{A}{B}{n} if and only if either
$\setsize{A}=\setsize{B}\leq i$ or $\setsize{A}<\setsize{B}$ with
$\setsize{A}\leq j$. 

In view of the discussion following Proposition \ref{p:sieber}, no
definability information can be inferred for two
functions with the same p-level. However, functions with different
p-levels yield immediate inexpressibility results:
\begin{cor}
\label{t:inexp}
Given $f$ and $g$ first-order monotone boolean functions
with p-levels of $(i_{f},j_{f})$ and $(i_{g},j_{g})$ respectively. If
$i_{f} > i_{g}$ or $j_{f} > j_{g}$, then $g \parnleq\ f$.
\end{cor}
\COMMENTOUT{
\begin{pf2}
Immediate by definition of p-levels, Proposition \ref{p:sieber} and
the Reduction and Closure lemmas.\qed
\end{pf2}
}

In summary, two integers are sufficient to completely characterize the 
set of presequentiality relations under which a function is
invariant. It turns out that these integers can be derived
straightforwardly from the trace of the function. Define the
\emph{coefficient of (linear) coherence} of a first-order monotone
boolean function $f$ by \[ \cc(f) =
\min\left\{\setsize{A}:A\subseteq\pi_{1}(\tr(f)), 
\setsize{A} \geq 2, \mbox{$A$ coherent} \right\} \]
with $\cc(f)$ defined to be $\infty$ when $\pi_{1}(\tr(f))$ has no
non-singleton linearly coherent subset. Similarly, define the 
\emph{bivalued coefficient of (linear) coherence} of a 
first-order monotone boolean function $f$ by 
\[ \bcc(f) = \min\left\{\setsize{A}:A\subseteq\pi_{1}(\tr(f)),
\setsize{A}\geq 3, \mbox{$A$ coherent and bivalued}\right\} \]
with $\bcc(f)$ is defined to be $\infty$ when $\pi_{1}(\tr(f))$ has no
non-singleton bivalued linearly coherent subset. We note that $\bcc(f)
\geq \cc(f)$ for all $f$.

The relationship between coefficients of coherence and
presequentiality levels is 
expressed by the following proposition, which provides a mechanical
way of determining the presequentiality level of a function, and hence
of determining the set of presequentiality relations under which a
function is invariant. 
\begin{lem}
\label{l:plevel}
Let \boolfn{f}{k} be a first-order monotone boolean function. Then $f$ has a p-level
of $(\bcc(f)-1,\cc(f)-1)$ (assuming standard rules for $\infty$). 
\COMMENTOUT{
where $\bcc(f)-1$ is $\infty$ when
$\bcc(f)=\infty$ and $\cc(f)-1$ is $\infty$ when $\cc(f)=\infty$.
}
\end{lem}
\begin{pf} 
We prove the result for $\cc(f)$. Consider the three cases:
\begin{enumerate}
\item \begin{sloppypar}
($\cc(f)=2$) We show that $f$ is invariant under
\preseq{\{1\}}{\{1,2\}}{2} but not \preseq{\{1,2\}}{\{1,2,3\}}{3}.
Assume $f$ is not invariant under \preseq{\{1\}}{\{1,2\}}{2}. Then
there exist tuples
$\left(x^{1}_{1},x^{1}_{2}\right),\cdots,\left(x^{k}_{1},x^{k}_{2}\right)\in \preseq{\{1\}}{\{1,2\}}{2}$
such that \tuplenotin{(y_{1},y_{2})}{\preseq{\{1\}}{\{1,2\}}{2}},
with $y_{i} = \applied{f}{x_{i}}{k}$. This means that $y_{1}\not=\bot$
and $y_{1}\not=y_{2}$. It is easy to see that
$(x^{1}_{1},\ldots,x^{k}_{1}) \leq (x^{1}_{2},\ldots,x^{k}_{2})$,
since for each $i\leq k$, either $x^{i}_{1}=\bot$ or
$x^{i}_{1}=x^{i}_{2}$. So by monotonicity of $f$, $y_{1}\leq
y_{2}$, contradicting $y_{1}\not=\bot$, and $y_{1}\not=y_{2}$. So $f$
must be invariant under \preseq{\{1\}}{\{1,2\}}{2}. On the other
hand, applying $f$ to the tuples
$\left(x^{1}_{1},x^{1}_{2},\bot\right),\cdots,\left(x^{k}_{1},x^{k}_{2},\bot\right) \in \preseq{\{1,2\}}{\{1,2,3\}}{3}$,
where the first two coordinates of the tuples are the elements of the first projection
of the trace forming a linearly coherent subset of size 2, yields the tuple $(\true,\true,\bot)$ or
$(\false,\false,\bot)$, neither of which is in
\preseq{\{1,2\}}{\{1,2,3\}}{3}.
      \end{sloppypar}

\item 
\begin{sloppypar}
($3\leq \cc(f)<\infty$) We show $f$ is invariant under
\preseqseq{\cc(f)-1}{\cc(f)}{\cc(f)} but not under
\preseqseq{\cc(f)}{\cc(f)+1}{\cc(f)+1}. Assume $f$ is not invariant under
\preseqseq{\cc(f)-1}{\cc(f)}{\cc(f)}. Then there exist tuples
\tuplearray{x}{\cc(f)}{k}{\preseqseq{\cc(f)-1}{\cc(f)}{\cc(f)}}
such that
\tuplenotin{\indexedtuple{y}{\cc(f)}}{\preseqseq{\cc(f)-1}{\cc(f)}{\cc(f)}}
with $y_{i}=\applied{f}{x_{i}}{k}$. This means that for all $i\leq
\cc(f)-1$, $y_{i}\not=\bot$ and there are $I,J$ with
$y_{I}\not=y_{J}$. Let $C\subseteq\pi_{1}(\tr(f))$ be an Egli-Milner
lowerbound of the first $\cc(f)-1$ coordinates of the given tuples,
$\setsize{C}\leq \cc(f)-1$. We cannot have
$\setsize{C}=1$ (say $C=\{v\}$), since that would imply that
$v\leq (x^{1}_{\cc(f)},\ldots,x^{k}_{\cc(f)})$: for each
$i\leq k$, either one of $x^{i}_{j}=\bot$ for $j\leq \cc(f)-1$ (hence $v_{j}=\bot$) or $x^{i}_{j}=x^{i}_{j'}$ for all
$j,j'\leq \cc(f)-1$ (hence $v_{j}\leq
x^{i}_{j}=x^{i}_{\cc(f)}$). But monotonicity of $f$ would imply that
for all $i,j$, $y_{i}=y_{j}$, a contradiction. Hence, $\setsize{C}\geq
2$. But since the first $\cc(f)-1$ coordinates of the given tuples form a
coherent subset, $C$ being an Egli-Milner lowerbound must also be
coherent (by a result in \cite{Bucciarelli97}). But this contradicts the fact that the minimal size for a
coherent subset of $\pi_{1}(\tr(f))$ is $\cc(f)$. So, $f$ is invariant
under \preseqseq{\cc(f)-1}{\cc(f)}{\cc(f)}. On the other hand, consider
the tuples
$\left(x^{1}_{1},\ldots,x^{1}_{\cc(f)},\bot\right),\cdots,\left(x^{k}_{1},\ldots,x^{k}_{\cc(f)},\bot\right)\in \preseqseq{\cc(f)}{\cc(f)+1}{\cc(f)+1}$
where the first $\cc(f)$ coordinates are the elements of a coherent subset 
of size $\cc(f)$ of $\pi_{1}(\tr(f))$ (which exists by
assumption). Appplying $f$ to these tuples yields a tuple
$(y_{1},\ldots,y_{\cc(f)},\bot)$ with $y_{i}\not=\bot$ for $i\leq
\cc(f)$, which cannot be in \preseqseq{\cc(f)}{\cc(f)+1}{\cc(f)+1}. 
\item ($\cc(f)=\infty$) We show that $f$ is invariant under all
presequentiality relations of the form \preseqseq{i}{i+1}{i+1}. Assume
that there exists an $i$ such that $f$ is not invariant under
\preseqseq{i}{i+1}{i+1}. The same reasoning as in the previous case
leads to a contradiction, although instead of contradicting the minimal size of
a coherent subset of $\pi_{1}(\tr(f))$ being $\cc(f)$, we contradict the
fact that there is no coherent subset of $\pi_{1}(\tr(f))$.
\end{sloppypar}
\end{enumerate}

The argument for $\bcc(f)$ is similar. \qed
\end{pf}

We can use Lemma \ref{l:plevel} to show that
presequentiality levels are preserved by the least upperbound
operation on functions in a natural way:
\begin{lem}
\label{l:plevelsup}
Given $f$ and $g$ first-order monotone boolean functions
with p-levels of $(i_{f},j_{f})$ and $(i_{g},j_{g})$
respectively. Then the p-level of $f+g$ is
\[(\min(i_{f},i_{g}),\min(j_{f},j_{g}))\].
\end{lem}
\begin{pf} 
Immediate by Lemma \ref{l:plevel} and the definition of $f+g$
in terms of $f$ and $g$. \qed
\end{pf}

It is not hard to check that any first-order monotone boolean
function has a p-level $(i,j)$ with $i\geq 2$ and $j\geq 1$ (consider 3
cases: $\cc(f)=\infty,cc(f)<\infty=\bcc(f), \bcc(f)<\infty$). We can
easily characterize sequential functions:

\begin{prop}
\label{l:continuous}
A first-order monotone boolean function has a
p-level of $(\infty,\infty)$ if and only if it is sequential
\end{prop}
\begin{pf}
($\Rightarrow$) It is sufficient to show that if $\cc(f)=\infty$, then 
$f$ is sequential. Let us first prove the following auxiliary result:
given \boolfn{f}{k+1} a monotone function and 
\boolfn{f'}{k} defined by \[ f'(x_{1},\ldots,x_{k})
= f(x_{1},\ldots,y,\ldots,x_{k}) \] for some fixed $y$ as the
$i^{\mbox{th}}$ argument of $f$. Then $\cc(f')\geq \cc(f)$.

Consider the two cases:
\begin{enumerate}
\item ($\cc(f)=\infty$) In this case, there is no linearly coherent
subset of $\pi_{1}(\tr(f))$, and hence there can be no linearly
coherent subset of $\pi_{1}(\tr(f'))$ (otherwise, it would
yield a linearly coherent subset of $\pi_{1}(\tr(f))$. Hence,
$\cc(f')=\infty \geq \cc(f)$ by definition.
\item ($\cc(f)<\infty$) Given $A\subseteq\pi_{1}(\tr(f'))$ a
coherent subset of size $\cc(f')$. Let $B$ be the following set:
\[ \left\{(x_{1},\ldots,x_{k+1})\in\pi_{1}(\tr(f)) :
(x_{1},\ldots,x_{i-1},x_{i+1},\ldots,x_{k+1})\in A, x_{i} \leq
y\right\}.\] We check that $B\subseteq\pi_{1}(\tr(f))$ is linearly
coherent. First, notice that $\setsize{B} = \setsize{A}$. Moreover, we
see that for all tuples in $B$, the $\mbox{i}^{\mbox{th}}$ position is
either a \bottom\ or a value $y$. Added to the fact that $A$ is
linearly coherent, we see that $B$ must be linearly coherent,
and hence $\cc(f)\leq \cc(f')$.
\end{enumerate}
And this proves the auxiliary result.

\COMMENTOUT{
($\Rightarrow$) We shall prove the contrapositive, namely that if
$\cc(f)<\infty$, then $f$ cannot be sequential.

We prove this by induction on the arity of $f$. First,
note that if $f$ has arity 1, it cannot have a linearly coherent
subset of the first projection of the trace. If $f$ has arity 2, it is
easy to see that $\pi_{1}(\tr(f))$ having a linearly coherent subset
implies that $f$ is not stable, and hence not sequential.

\textbf{(induction step)} Let $f$ be a function of arity $k+1$. Let
$A\subseteq\pi_{1}(\tr(f))$, A coherent, 
$\setsize{A}\geq 3$ (since $f$ is stable). We consider two cases:
\begin{enumerate}
\item A is \bottom-covering. Then it is easy to see that $f$ cannot be
sequential (no index of sequentiality).
\item There exists an index of sequentiality $j$. Let $y$ be the value
of the tuples of $A$ at position $j$. Consider the function
$f' = f(x_{1},\ldots,y,\ldots,x_{k+1})$ of arity $k$. The
trace of this function must contain the tuples of $A$ (minus column
$j$), and these form a coherent subset of
$\pi_{1}(\tr(f'))$. So by induction hypothesis, $f'$ is
not sequential, and hence neither is $f$.
\end{enumerate}
}

We now prove the sufficient condition by induction on the arity of $f$.

\textbf{(base case)} \function{f}{\bool}{\bool}. Consider $f(\bot)$. If
$f(\bot)\not=\bot$, then by monotonicity $f$ is constant, and hence
sequential. if $f(\bot)=\bot$, then consider $f(y)$ for a fixed
$y$. This must be a constant, so $f$ is sequential (by the definition
of sequentiality). 

\textbf{(induction step)} Assume the result holds for all functions of arity
$k$. Consider \boolfn{f}{k+1}, with $\cc(f)=\infty$.
\begin{enumerate}
\item We first need to show that there exists an index of
sequentiality. Assume not: for all $i$, for any fixed
$x_{j}, \forall j\not=i$,
$f(x_{1},\ldots,\bottom,\ldots,x_{k+1})\not=\bot$. Then
$\pi_{1}(\tr(f))$ must be \bottom-covering, which contradicts
$\cc(f)=\infty$. 
\item Given $i$ the index of sequentiality of $f$, look at the
function $f'(z_{1},\ldots,z_{k}) =
f(z_{1},\ldots,y,\ldots,z_{k})$ for a fixed $y$ in position $i$. By
the auxiliary result, $\cc(f')=\infty$, and the induction
hypothesis applies to show that $f'$ and therefore $f$ must be
sequential.\qed
\end{enumerate}

($\Leftarrow$) Immediate, since $f$ sequential implies that $f$ is
PCF-definable, and hence $f$ must be invariant under all sequentiality
relations --- including presequentiality relations.

\end{pf}

\section{Structural results}
\label{s:struct}

In this section, we use p-levels to guide our exploration of the
\CONT{} semilattice. The approach is roughly as follows: we identify
interesting classes of functions (stable functions, unstable
functions, stable-dominating functions, monovalued functions), and
show that they have a natural characterization in 
terms of p-levels. We then use the p-level characterization to look
for interesting natural hierarchies. A hierarchy is deemed natural if
it is made up of functions living on different p-levels. We
also show that interesting well-known functions also have a natural
characterization in terms of p-levels.

\subsection{The \STABLE{} semilattice}

Define a \emph{stable degree of parallelism} to be a degree of parallelism
containing at least one stable function. We can characterize stable degrees in terms of
p-levels:
\begin{prop}
\label{p:stable}
A degree of parallelism is stable if and only if its 
p-level is of the form $(i,j)$ with $i\geq 2$ and $j\geq 2$
\end{prop}
\begin{pf}
($\Rightarrow$) Given $f$ a stable function. Then $\cc(f)\geq 3$,and by
Lemma \ref{l:plevel}, $f$ must have a p-level of the form
$(i,j)$ with $j\geq \cc(f)-1 \geq 2$. Since $f$ is monotone, $i\geq 2$.

($\Leftarrow$) Given $f$ with a p-level $(i,j)$ with $j\geq 2$. By
Lemma \ref{l:plevel}, $\cc(f)-1 \geq 2$, so that $\cc(f)\geq
3$. Hence, $f$ must be stable. \qed
\end{pf}

As a consequence, every function in a stable degree of parallelism must be
stable. Let \STABLE{} be the 
subposet of \CONT{} consisting of all stable degrees of parallelism. 
\begin{prop}
\STABLE{} is a subsemilattice of \CONT.
\end{prop}
\begin{pf} 
It is easy to see that the least upperbound of two stable degrees of
parallelism is itself a stable degree of parallelism. The degree of
sequential functions is the bottom element of the semilattice and the
Berry-Plotkin function (\funbp) is its top element, as noted by Plotkin
and reported by Curien in \cite{Curien93}. \qed
\end{pf}

The Berry-Plotkin function is defined by the following trace: 
\begin{center}
\begin{tabular}{|ccc|c|} \hline
\bottom & \true & \false & \true \\
\true & \false & \bottom & \false \\
\false & \bottom & \true & \false \\ \hline
\end{tabular}
\end{center}

We can in fact completely characterize the degree of parallelism of \funbp\ via
presequentiality levels:
\begin{prop}
Given $f$ a first-order monotone boolean function. Then $f$ has a p-level of
$(2,2)$ iff $f \pareq \funbp$.
\end{prop}
\begin{pf}
($\Rightarrow$) Given $f$ with a p-level of $(2,2)$. This means that
$\bcc(f) = 3$, in other words, there exists an $A\subseteq\pi_{1}(\tr(f))$
bivalued and linearly coherent, with $\setsize{A}=3$. We can assume
without loss of generality that one element of $A$ returns \true\ and
the remaining two return \false\ (otherwise, consider $\fneg(f)$ which
is equiparallel to $f$ and has the desired property). Define
\function{g}{\tr(\funbp)}{\tr(f)} by sending the first trace element of
\funbp\ 
(the one returning \true) to the element of $A$ returning \true, and
the remaining elements of \funbp\ to the elements of $A$ returning
\false. Since $A$ is linearly coherent, it is clear that $g$ satisfies
the condition of Proposition \ref{p:buccmala}, and $\funbp\parleq
f$, Hence by Proposition \ref{p:stable}, $f$ is stable, so $f\parleq\funbp$.

($\Leftarrow$) Given $f \pareq\funbp$. Then $f$ must be invariant under
the same sequentiality relations, hence the p-level of $f$ is the same
as the p-level of \funbp, namely $(2,2)$.\qed

\end{pf}

\subsection{The Gustave hierarchy}

The structure of \STABLE{} is non-trivial. Since the functions
\bucc{n}{m} are easily seen to be stable, the whole Bucciarelli
hierarchy is in \STABLE. We can identify a subhierarchy of the
Bucciarelli hierarchy derived from the Gustave function \cite{Berry76}. The Gustave
function \gust{} is given by the following trace (in
matrix form):
\begin{center}
\begin{tabular}{|ccc|c|} \hline
\bottom & \true & \false & \true \\
\true & \false & \bottom & \true \\
\false & \bottom & \true & \true \\ \hline
\end{tabular}
\end{center}

\begin{defn}
Let \boolfn{\gust{i}}{2i+1} ($i\geq 1$) be defined by the following trace
(in matrix form): 
\begin{center}
\begin{tabular}{|cccccc|c|} \hline
\bottom & \true & \false & $\cdots$ & \true & \false & \true \\
\false & \bottom & \true & $\cdots$ & \false & \true & \true \\
\true & \false & \bottom & $\cdots$ & \true & \false & \true \\
& $\vdots$ & & & & & $\vdots$ \\
\false & \true & \false & $\cdots$ &  \bottom & \true & \true \\
\true & \false & \true & $\cdots$ & \false & \bottom & \true \\ \hline
\end{tabular}
\end{center}
\end{defn}

Note that \gust{1} is just \gust{}. It is easy to verify the
following: 
\begin{prop}
$\gust{i} \pareq \bucc{2i+1}{2i+1}$.
\end{prop}
\begin{pf} 
First note that a monovalued first-order monotone boolean function with
$\setsize{\tr(f)}=\cc(f)=n$ is such that $f\pareq \bucc{n}{n}$, by an
application of Proposition \ref{p:buccmala}, and note
that $\setsize{\tr(\gust{i})}=\cc(\gust{i})=2i+1$. \qed
\end{pf}

By Lemma \ref{l:plevel}, the functions \gust{i} have a p-level of $(\infty,2i)$. This characterization
allows us to derive the following result:
\begin{prop}
\label{p:nominimal}
There is no minimal stable non-sequential function.
\end{prop}
\begin{pf} 
Assume $g$ is a stable non-sequential function that is
minimal, i.e. for all $f$, $f$ stable, non-sequential, $g \parleq f$.

Since $g$ is not sequential, by Proposition \ref{l:continuous}, there must be
some $A,B,n$ such that $g$ is not invariant under \preseq{A}{B}{n}. 

Consider \gust{\setsize{A}}. By the p-level of \gust{i} functions, since
$\setsize{A}\leq 2\setsize{A}$, \gust{\setsize{A}} is invariant under
\preseq{A}{B}{n}. 

Hence $g \parnleq \gust{\setsize{A}}$, a contradiction. \qed
\end{pf}

On the other hand, we can show that the Gustave hierarchy is co-final
in the non-sequential functions, that is any non-sequential function
must dominate one of the functions in the hierarchy. 
\begin{prop}
\label{p:cofinal}
Given $f$ a stable non-sequential first-order monotone boolean function. Then there
exists an integer $i$ such that $\gust{i} \parleq f$.
\end{prop}
\begin{pf} 
The function $f$ being non-sequential implies that $\cc(f)<\infty$ by
Propositions \ref{l:continuous} and \ref{l:plevel}. Moreover, $f$ being
stable implies that  
$\cc(f)\geq 3$ (by Lemma \ref{l:plevel} and Proposition
\ref{p:stable}). Let $A$ be a linearly coherent 
subset of 
$\pi_{1}(\tr(f))$ of size $\cc(f)$. Define a arbitrary function
\function{g}{\tr(\gust{\cc(f)})}{\tr(f)} with $\pi_1(g(\tr(\gust{\cc(f)})))=A$. It is easy to
see that the conditions of Proposition \ref{p:buccmala} are
satisfied, so that $\gust{\cc(f)}\parleq f$. \qed
\end{pf}

Note that Propositions \ref{p:nominimal} and \ref{p:cofinal} can be
derived directly from Bucciarelli's result. We 
merely identify a natural subset of the Bucciarelli hierarchy that is
sufficient for our purpose.

\subsection{The Bivalued-Gustave hierarchy}

Functions in the Gustave hierarchy (and indeed, in the
Bucciarelli hierarchy) are all monovalued. We return to monovalued
functions in Section \ref{s:mono}. For now, let us extend the 
Gustave hierarchy to a hierarchy of bivalued functions, the Bivalued-Gustave
hierarchy.

\begin{defn}
Let \boolfn{\bgust{i}{j}}{2i+1} $(j\leq i)$ be the function defined by
the following trace (in matrix form):
\begin{center}
\begin{tabular}{|cccccc|c|} \hline
\bottom & \true & \false & $\cdots$ & \true & \false & $r_{1}$ \\
\false & \bottom & \true & $\cdots$ & \false & \true & $r_{2}$ \\
\true & \false & \bottom & $\cdots$ & \true & \false & $r_{3}$ \\
& $\vdots$ & & & & & $\vdots$ \\
\false & \true & \false & $\cdots$ &  \bottom & \true & $r_{2i}$ \\
\true & \false & \true & $\cdots$ & \false & \bottom & $r_{2i+1}$ \\ \hline
\end{tabular}
\end{center}
with \[ r_{l} = \left\{ \begin{array}{ll}
                         \false & \mbox{if $1\leq l\leq j$} \\
                         \true & \mbox{otherwise} \\
                        \end{array} \right. \]
\end{defn}

Let us first show that the $j$ parameter in $\bgust{i}{j}$ is
unnecessary: we may pick \bgust{i}{1} as a representative of the class 
of \bgust{i}{j} functions, and drop the superscript to refer to the function
as \bgustf{i}.

\begin{lem}
\label{l:bg2}
Given $j,j' \leq i$,
$\bgust{i}{j}\pareq\bgust{i}{j'}$.
\end{lem}
\begin{pf} 
We prove by induction on $j$ that for all $
j$,$\bgust{i}{j}\pareq\bgust{i}{1}$. The case $j=1$ is trivial. For the
induction step ($j\geq 2$), assume that
$\bgust{i}{j-1}\pareq\bgust{i}{1}$ and consider \bgust{i}{j}. We show
$\bgust{i}{j}\pareq\bgust{i}{j-1}$. Define the following terms:
\begin{eqnarray*}
M_{1} & = & \lambda f\lambda x_{1}\ldots x_{2i+1}.\mbox{if
$f(x_{1},\ldots,x_{2i+1})$} \\
& & \mbox{then $f(x_{2},\ldots,x_{2i+1},x_{1})$ else \false\ fi} \\
M_{2} & = & \lambda f\lambda x_{1}\ldots x_{2i+1}.\mbox{if
$f(x_{1},\ldots,x_{2i+1})$} \\
& & \mbox{then \true\ else $f(x_{2i+1},x_{1},\ldots,x_{2i})$ fi}
\end{eqnarray*}
It is not hard to see that $\bgust{i}{j} =
\Mean{M_{1}}\bgust{i}{j-1}$ and $\bgust{i}{j-1} =
\Mean{M_{2}}\bgust{i}{j}$, thereby showing
$\bgust{i}{j}\pareq\bgust{i}{j-1}\pareq\bgust{i}{1}$ by the induction
hypothesis. \qed
\end{pf}

It remains to show that the functions \bgustf{i} actually form a hierarchy.
First note that by Lemma \ref{l:plevel} \bgustf{i} has a p-level of $(2i,2i)$. 
\begin{prop}
\label{p:bghier}
$\bgustf{i}\parleq\bgustf{j}$ iff $i\geq j$. 
\end{prop}
\begin{pf} 
($\Leftarrow$) A straightforward application of Proposition
\ref{p:buccmala}: consider any surjective function
\function{g}{\tr(\bgustf{i})}{\tr(\bgustf{j})} sending the unique trace element
returning \true\ to the unique trace element returning \true, and any trace
element returning \false\ to any trace element returning \false. It is
easy to see that all conditions of Proposition \ref{p:buccmala}
are satisfied, and $\bgustf{i}\parleq\bgustf{j}$.

$(\Rightarrow$) Assume $i<j$. The p-level of \bgustf{i} is $(2i,2i)$
and the p-level of \bgustf{j} is $(2j,2j)$. By Corollary \ref{t:inexp},
$\bgustf{i}\parnleq\bgustf{j}$. \qed
\end{pf}

The following result is immediate:
\begin{prop}
\label{prop:4.11}
For all $i$, $\gust{i}\parleq\bgustf{i}$.
\end{prop}
\begin{pf} 
Via Proposition \ref{p:buccmala}. \qed
\end{pf}

Combining functions in the Gustave hierarchy and the Bivalued-Gustave
hierarchy via the least upperbound operation produces a two-dimensional
hierarchy, with functions of the form $\bgustf{i} +
\gust{j}$. A trivial application of Lemma \ref{l:plevelsup} gives
a p-level of $(2i,2\min(i,j))$ for $\bgustf{i} + \gust{j}$. This allows
us to derive the following governing equations describing the
structure of the hierarchy: 
\begin{prop}
\label{prop:4.12}
$\bgustf{i}+\gust{j}\parleq \bgustf{i'}+\gust{j'}$ iff
$i'\leq i$ and $\min(i',j')\leq\min(i,j)$. 
\end{prop}
\begin{pf}
$(\Rightarrow)$ We prove the contrapositive. If $i < i'$ or $\min(i,j) <
\min(i',j')$, then by Corollary \ref{t:inexp} and the
p-level of functions in the 
hierarchy, $\bgustf{i}+\gust{j}\parnleq\bgustf{i'}+\gust{j'}$.

$(\Leftarrow)$ Since $i'\leq i$, Proposition
\ref{p:bghier} tells us that
$\bgustf{i}\leq\bgustf{i'}\leq\bgustf{i'}+\gust{j'}$.
We then consider three cases:
\begin{enumerate}
\item ($\min(i,j)=i$) Proposition \ref{prop:4.11} implies that
\[\gust{j}\parleq\bgustf{j}\parleq\bgustf{i}\parleq\bgustf{i'}+\gust{j'}\]
Hence,
$\bgustf{i}+\gust{j}\parleq\bgustf{i'}+\gust{j'}$.
\item ($\min(i,j)=j$, $\min(i',j')=i'$) By
assumption, $i'\leq j$, and hence by Proposition
\ref{prop:4.11},
$\gust{j}\parleq\bgustf{j}\parleq\bgustf{i'}\parleq\bgustf{i'}+\gust{j'}$.
Hence $\bgustf{i}+\gust{j}\parleq\bgustf{i'}+\gust{j'}$.
\item ($\min(i,j)=j$, $\min(i',j')=j'$) By
assumption, $j'\leq j$, and hence 
\[\gust{j}\parleq\gust{j'}\parleq\bgustf{i'}+\gust{j'}\]
Hence $\bgustf{i}+\gust{j}\parleq\bgustf{i'}+\gust{j'}$. \qed
\end{enumerate}

\end{pf}

\subsection{The \UNSTABLE{} semilattice}

Define an \emph{unstable degree of parallelism} to be a degree of parallelism
containing no stable function. It is easy to show that a degree of
parallelism is unstable if and only if it has a p-level of the form $(i,1)$
with $i\geq 2$, by Proposition \ref{p:stable}. Let \UNSTABLE{} be the
subposet of \CONT{} consisting of all 
unstable degrees of parallelism. Define the Detector function (\fundet) to simply return \true\ if
one of its two inputs has a value (\true\ or \false\
indifferently). For various reasons, it is simpler to work with the
following function \funtdet\, which is easily seem to be equiparallel
to \fundet: 
\begin{center}
\begin{tabular}{|cc|c|} \hline
\true & \bottom & \true \\
\bottom & \true & \true \\ \hline
\end{tabular}
\end{center}

\begin{prop}
\UNSTABLE{} is a subsemilattice of \CONT.
\end{prop}
\begin{pf}
It is easy to see that the least upperbound of two unstable degrees of
parallelism is unstable. The top element of \UNSTABLE{} is the degree of \funpor\
and its bottom element is the degree of the Detector function. This
last fact is an application of Proposition
\ref{p:buccmala}: given $f$ an unstable first-order monotone boolean
function; since $f$ is unstable, there must exist $A\subseteq\pi_{1}(\tr(f))$ with $A$
coherent and $\setsize{A}=2$. Define a function
\[ \function{g}{\tr(\funtdet)}{\tr(f)} \] with the only constraint that
each element of the trace of \funtdet\ goes to a distinct element of
the trace of $f$ corresponding to the subset $A$. It is easy to see
that all the conditions of Proposition \ref{p:buccmala} are met,
hence $\funtdet\parleq f$.  \qed
\end{pf}

Detector first appeared in the context of asynchronous dataflow
networks. Rabinovich shows in \cite{Rabinovich98} that \fundet\
is minimal among unstable functions in that context.

A degree of parallelism is unstable if and only if it is not stable, so we
see that the \STABLE{} and the \UNSTABLE{} semilattices form a partition
of the full \CONT{} semilattice. We presently identify one hierarchy
of functions in \UNSTABLE{} (another will be presented in Section \ref{s:sdom}
); functions in this hierarchy are derived from \funpor:
\begin{defn}
Let \boolfn{\por{i}}{i} ($i\geq 2$) be defined by the following trace
(in matrix form): 
\begin{center}
\begin{tabular}{|cccccc|c|} \hline
\true & \true & \true & $\cdots$ & \true & \bottom & \true \\
\true & \true & \true & $\cdots$ & \bottom & \true & \true \\
& $\vdots$ & & & & & $\vdots$ \\
\true & \true & \bottom & $\cdots$ & \true & \true & \true \\
\true & \bottom & \true & $\cdots$ & \true & \true & \true \\
\bottom & \true & \true & $\cdots$ & \true & \true & \true \\
\false & \false & \false & $\cdots$ & \false & \false & \false \\ \hline
\end{tabular}
\end{center}
\end{defn}

Note that \por{2} is just \funpor. \por{i} takes $i$
inputs and returns \true\ if at least $i-1$ are \true, and \false\ if
all are \false. These functions span the whole range of
allowable p-levels for unstable functions as the next proposition
shows: 
\begin{prop}
\label{p:porplevel}
\por{i} has a p-level of $(i,1)$.
\end{prop}
\begin{pf} 
Since \por{i} is monotone and unstable, it must have a
p-level of the form $(j,1)$ for some $j\geq2$, by the characterization
of p-levels of monotone and stable functions.

By inspection, we see that the only bivalued coherent subset of
$\pi_{1}(\tr(\por{i}))$ is $\pi_{1}(\tr(\por{i}))$ itself. Hence,
$\bcc(f)=i+1$ and by Lemma \ref{l:plevel}, $j= \bcc(f)-1
= i$. \qed
\end{pf}

These functions indeed form a hierarchy:
\begin{prop}
\label{prop:5.4}
$\por{i}\parleq \por{j}$ iff $i\geq j$. 
\end{prop}
\begin{pf} 
($\Leftarrow$) Consider the following PCF-term:
\[ M = \lambda f.\lambda x_{1}\ldots
x_{i+1}.ALLEQ(t_{1}(x_{1},\ldots,x_{i+1}),\ldots,t_{i+1}(x_{1},\ldots,x_{i+1}))
\]
where
\[ ALLEQ = \lambda x_{1}\ldots x_{i+1}.\mbox{if $(x_{1}=\ldots=x_{i+1})$ then
$x_{1}$ else $\bot$ fi} \]
which returns the value $v$ if and only if all the arguments have the
value $v$.

Each $t_{j}$ is an application of \por{i} to a subset of $i$ inputs
out of the $i+1$ possible inputs. Since $\choosefn{i+1}{i} = i+1$, there
are $i+1$ such terms. We claim this term is such that $\por{i+1} =
\Mean{M}\por{i}$. 

\begin{enumerate}
\item The $t_{j}$ functions all return \true\ iff at least $i$ \true's
appear in their arguments
\begin{enumerate}
\item (at least $i$ \true's) Each subset of size $i$ has at least
$i+1$ \true's, so each $t_{j}$ function returns \true.
\item (less then $i$ \true's) There exists one subset of size $i$ with
less than $i-1$ \true's, so the corresponding $t_{j}$ function returns
\bottom.
\end{enumerate}

\item The $t_{j}$ functions all return \false\ iff all inputs are \false.
\begin{enumerate}
\item (all \false's) Every $t_{j}$ returns \false.
\item (not all \false's) There exists a subset of size $i$ with not
all inputs being \false. The corresponding $t_{j}$ does not return
\false. 
\end{enumerate}
\end{enumerate}

($\Rightarrow$) Assume $i<j$. The result is immediate by Corollary
\ref{t:inexp} and Proposition \ref{p:porplevel}. 
\end{pf}

\subsection{The \SDOM{} semilattice}
\label{s:sdom}

It is clear that unstable functions are strictly more powerful than
stable functions, in the sense that no stable function can implement
an unstable function, but unstable functions can implement stable
functions. In this
section, we characterize the unstable functions that can
implement all stable functions, and show that they form a
subsemilattice of \UNSTABLE. 

\begin{defn}
Let $f$ be an unstable first-order monotone boolean function. We say $f$ is
\emph{stable-dominating} if for any stable first-order monotone
boolean function $g$, we have $g \parleq f$.
\end{defn}

Since the \STABLE{} semilattice has a top element \funbp, a necessary and
sufficient condition for an unstable function $f$ to be stable-dominating is
to have $\funbp\ \parleq f$. Since any stable-dominating function must also
dominate \fundet\ (the bottom element of \UNSTABLE), we have that $f$ is
stable-dominating if and only if $\funbp+\fundet\ \parleq f$. This allows us to
derive the following characterization of stable-dominating functions:
\begin{prop}
\label{p:sdomchar}
Given $f$ an unstable first-order monotone boolean function. Then $f$ is
stable-dominating iff $f$ has a p-level of $(2,1)$.
\end{prop}
\begin{pf}
($\Rightarrow$) Assume $f$ is stable-dominating. Then by previous
argument, $\funbp+\fundet\parleq f$. Since \funbp\ has p-level $(2,2)$ and
\fundet\ has p-level $(\infty,1)$, $\funbp+\fundet$ has p-level $(2,1)$
by Lemma \ref{l:plevelsup}. Assume $f$ does not have a p-level of
$(2,1)$. By Proposition \ref{l:continuous}, $f$ must have a p-level of
$(i,j)$ with $i\geq 2$, $j\geq 1$ and $i\not=2$ or 
$j\not=1$. But by Corollary \ref{t:inexp}, we get that
$\funbp+\fundet\parnleq f$, a contradiction.

($\Leftarrow$) Given $f$ with p-level $(2,1)$. By the characterization
of the p-level of stable functions, $f$ is unstable. We need only
check that 
$\funbp\parleq f$. By Lemma \ref{l:plevel}, $\bcc(f)=3$. Let $A$
be the subset of $\pi_{1}(\tr(f))$ of size 3. Assume without loss of
generality that $A$ has one element returning \true\ and two elements
returning \false\ (if not, consider $\fneg(f)$ which is equiparallel to
$f$). Define a function \function{g}{\tr(\funbp)}{\tr(f)} sending the
element of the trace of \funbp\ returning \true\ to the element of $A$
returning \true\ and the elements of the trace of \funbp\ returning \false\
to the elements of $A$ returning \false. It is easy to see that all
the conditions of Proposition \ref{p:buccmala} hold, and hence we
have $\funbp\parleq f$. So $f$ is stable-dominating. \qed
\end{pf}

Define a \emph{stable-dominating degree of parallelism} to be a degree of
parallelism containing a stable-dominating function. By Proposition
\ref{p:sdomchar}, every
function in a stable-dominating degree of parallelism is
stable-dominating. Let \SDOM{} be the subposet of \CONT{} (in fact, of
\UNSTABLE) consisting of all stable-dominating degrees of
parallelism. 
\begin{prop}
\SDOM{} is a subsemilattice of \UNSTABLE. 
\end{prop}
\begin{pf}
It is easy to see by the above characterization that the least
upperbound of two stable-dominating degrees of parallelism is itself
stable-dominating. The bottom element of \SDOM{} is the degree of
$\funbp+\fundet$, and its top element is the degree of \funpor. \qed
\end{pf}

To show this subsemilattice is non-trivial, we exhibit an hierarchy
of functions in \SDOM{}. Note however that because stable-dominating
functions are all in the same p-level, we cannot show inexpressibility
using presequentiality relations. Consider the functions
$\funbp+\por{i}$, which are easily seen 
to be stable-dominating. Note that $\funbp+\por{2} \pareq \por{2} \pareq
\funpor$. These functions form a hierarchy:
\begin{prop}
\label{prop:6.4}
$\funbp+\por{i}\parleq\funbp+\por{j}$ iff $i\geq j$. 
\end{prop}
\begin{pf}
($\Leftarrow$) We know $\funbp\parleq\funbp+\por{j}$ for all
$j\geq 2$. Similarly, by Proposition \ref{prop:5.4},
$\por{i}\parleq\por{j}\parleq\funbp+\por{j}$. Hence, by the property
of least upperbounds, we get that
$\funbp+\por{i}\parleq\funbp+\por{j}$. 

($\Rightarrow$) Assume $i<j$. Define the following sequentiality
relation of arity $j$ 
\[ R =
\preseq{\{1,2\}}{\{1,2\}}{j}\cap\cdots\cap\preseqseq{j}{j}{j}
\]
By Proposition \ref{p:sieber}, it is sufficient to show that $\funbp+\por{j}$
is invariant under $R$, but $\funbp+\por{i}$ is not.
\begin{enumerate}
\item \begin{sloppypar}
($\funbp+\por{j}$ invariant) Going back to the definition of
$+$, without loss of generality we can take 
\[(\funbp+\por{j})(\true,x_{1},\ldots,x_{j})= \por{j}(x_{1},\ldots,x_{j})\]
For the sake of contradiction, assume $\funbp+\por{j}$ is not invariant under
$R$. Then there exists tuples
\tuplearray{x}{j}{k}{R}
\COMMENTOUT{with $k=j$ if $i\geq 3$ and $k=4$ for $i=2$}. Let $y$ =
$\indexedtuple{y}{j}$, with $y_{m}=\applied{\funbp+\por{j}}{x_{m}}{k}$,  and
$y\not\in R$. 
      \end{sloppypar}

By induction on $2\leq m\leq j$, we show $\funbp+\por{j}$ must be invariant
under \preseqseq{m}{m}{j}. For $m=2$, 
$\funbp+\por{j}$ is invariant under
\preseq{\{1,2\}}{\{1,2\}}{j} by the Closure Lemma and Proposition
\ref{l:continuous}.

For the induction step, assume for the sake of contradiction that $\funbp+\por{j}$ is not
invariant under \preseqseq{m+1}{m+1}{j}. Then there is no \bottom\
in $y_{1},\ldots,y_{m+1}$, and there exists $I,J$ with
$y_{I}\not=y_{J}$. By the induction hypothesis, $\funbp+\por{j}$ is
invariant under \preseqseq{m}{m}{j}, so we must have
$y_{1}=\cdots=y_{m}$, and hence the only possibility is that
$y_{m+1}\not=y_{1}$. Since no \bottom\ appears in the resulting tuple,
the first tuple above must all be \true\ or all be \false, by the
definition of $+$. If it is
all \false, then the columns of the tuples must come from the trace of \funbp, but
since the first $m$ columns are linearly coherent and return the same
result, this would mean that the Egli-Milner lowerbound of the first
$m$ column has only one element, and since it is also coherent with
the last column (which returns a different result), this contradicts
\funbp\ being stable. Hence, the first tuple must be all \true, and
the columns must come from the trace of \por{j}. But the $m+1$
columns form a linearly coherent set of size less than or equal to
$j$, and we can easily show that they cannot contain the trace element of $\por{j}$
that returns false. So we must have $y_{m+1} = y_1$. 

Therefore, $\funbp+\por{j}$ is invariant under
\preseqseq{m}{m}{j} for $2\leq m\leq j$, hence
$\funbp+\por{j}$ is invariant under $R$.

\item ($\funbp+\por{i}$ not invariant) Again without loss of
generality, we can take
\[(\funbp+\por{i})(\true,x_{1},\ldots,x_{i})=\por{i}(x_{1},\ldots,x_{i})\]
We show that $\funbp+\por{i}$ is not invariant under
\preseqseq{i+1}{i+1}{j}, implying it is not invariant under $R$. 
Consider the following tuples of length $j$:
\[\left( \true\ldots,\true\right),\left(
x_1^1,\ldots,x_{i+1}^1,\bot,\ldots,\bot\right),\cdots,
\left(x_1^{i},\ldots,x_{i+1}^i,\bot,\ldots,\bot\right)\] where
$\{(\true,x_{m}^1,\ldots,x_m^i)\}$ ($m\leq i+1$) is the subset of the first projection
of the trace of $\funbp+\por{i}$ corresponding to $\por{i}$. It is
easy to see that all those tuples are in
\preseqseq{i+1}{i+1}{j}. Applying $\funbp+\por{i}$ to the columns of
the tuples yields the tuple
$(\underbrace{\true,\ldots,\true}_i,\false,\bot,\ldots,\bot)$, which is
not in \preseqseq{i+1}{i+1}{j}. \qed

\end{enumerate} 
\end{pf}

\subsection{The \MONO{} semilattice}
\label{s:mono}

Up to this point all the semilattices we have introduced were related
in some way to the partitioning of
functions according to whether or not they were stable. We now consider
a different characteristic and derive a corresponding semilattice. Define a
\emph{monovalued degree of parallelism} to be a degree of parallelism
containing at least one monovalued function. We can characterize monovalued
degrees of parallelism by their p-level: 
\begin{prop}
\label{prop:7.1}
A degree of parallelism is monovalued if and only if its p-level is of
the form $(\infty,j)$ with $j\geq 1$.
\end{prop}
\begin{pf}
If $f$ is monovalued then $\bcc(f)=\infty$, since there can be no
bivalued coherent subset of $\pi_{1}(\tr(f))$. Moreover, since $f$ is
monotone, it must have a p-level of the form $(i,j)$ with $i\geq 2$
and $j\geq 1$. We know $i=\infty$ (since $\bcc(f)=\infty$), so $f$ must
have a p-level of the form $(\infty,j)$ with $j\geq 1$. \qed
\end{pf}

Let \MONO{} be the subposet of \CONT{} containing all monovalued
degrees of parallelism.
\begin{prop}
\MONO{} is a subsemilattice of \CONT.
\end{prop}
\begin{pf}
The least upperbound of two monovalued degrees of parallelism is itself
monovalued. The bottom element of \MONO{} is the degree of all sequential
functions, and its top element is the degree of \fundet, the Detector
function. To show this, consider $f$ a monovalued first-order monotone 
boolean function. 
Without loss of generality, assume $f$ always returns \true\ (if not,
consider $\fneg(f)$ which is equiparallel to $f$). Let \funntdet{n} be
the function of arity $n$ that returns \true\ if one of its arguments
is \true. It is not hard to show that for all $n$, $\funntdet{n}
\parleq \funtdet$. Let $n=\setsize{\tr(f)}$. Consider the following 
PCF-term: 
\[ M = \lambda p\lambda x_{1}\ldots
x_{k}. p(t_{1}(x_{1},\ldots,x_{k}),\ldots,t_{n}(x_{1},\ldots,x_{k}))
\]
where $t_{j}$ is a term checking if its arguments agree with the
$\mbox{j}^{\mbox{th}}$ element of $\pi_{1}(\tr(f))$ --- and returning
\true\ if they do and blocking if they don't. For example, for the
Gustave function \gust{}, the terms look like:
\begin{eqnarray*}
t_{1} & = & \lambda x_{1}x_{2}x_{3}.(x_{2} \wedge \neg x_{3}) \\
t_{2} & = & \lambda x_{1}x_{2}x_{3}.(x_{1} \wedge \neg x_{2}) \\
t_{3} & = & \lambda x_{1}x_{2}x_{3}.(x_{3} \wedge \neg x_{1})
\end{eqnarray*}
It is easy to see that $f = \Mean{M}\funntdet{n}$, and since
$\funntdet{n}\parleq\funtdet$, $f \parleq\funtdet$. \qed
\end{pf}

We note that the \MONO{} semilattice contains the
Bucciarelli hierarchy.

We can fully characterize the degree of parallelism of \fundet\ via
p-levels, as we did with \funbp:
\begin{prop}
Given $f$ a first-order monotone boolean function. Then $f$ has a p-level of
$(\infty,1)$ iff $f \pareq \fundet$.
\end{prop}
\begin{pf}
($\Rightarrow$) If $f$ has a p-level of $(\infty,1)$, then $f$ must be
both monovalued and unstable. By minimality of \fundet\ in the
\UNSTABLE{} semilattice, $\fundet\parleq f$.  Since \fundet\ is the top
element for monovalued functions and $f$ monovalued, $f \parleq
\fundet$. Hence $f \pareq\fundet$. 

($\Leftarrow$)  Given $f \pareq\fundet$. Then $f$ must be invariant under
the same sequentiality relations, hence the p-level of $f$ is the same
as the p-level of \fundet, namely $(\infty,1)$. \qed
\end{pf}

Since a function is unstable if and only if its p-level is $(i,1)$ for 
some $i \geq 2$, and it is monovalued if and only if its p-level is 
$(\infty,j)$ for some $j\geq 1$, $[\fundet]$ is the only unstable and
monovalued degree of parallelism.

We will mention a final interesting result concerning monovalued
degrees of parallelism. We can further characterize monovalued degrees
of parallelism, a notion involving the description of a function, via
extensional properties of the corresponding functions. A function $f$
is \emph{subsequential} if there exists a sequential function $g$ that 
extends $f$, that is that dominates $f$ in the extensional ordering on 
$\boolm{k}$.
\begin{prop}
A function $f$ is subsequential if and only if $[f]$ is monovalued.
\end{prop}
\begin{pf}
The proof is a corollary of the proposition in
\cite{UNSTABLE:Bucciarelli97} which in our terminology states that
given $f$ a first-order monotone boolean function, then $f$ is subsequential
iff $\bcc(f)=\infty$. 
By this proposition, $f$ is subsequential iff
$\bcc(f)=\infty$. By Lemma \ref{l:plevel}, $f$ is
subsequential iff $f$ has p-level $(\infty,j)$ for some $j\geq 1$. By
Proposition \ref{prop:7.1}, $f$ is subsequential iff $[f]$ is
monovalued.  \qed
\end{pf}

Therefore, every subsequential function is expressible by \fundet\ and
conversely, \fundet\ can only express subsequential functions.

\section{Conclusion}

In this paper, we set out to explore the structure of \CONT, the
semilattice of degrees of parallelism of first-order monotone
boolean functions. It is known that Sieber's sequentiality relations
fully characterize the ordering on the semilattice. By turning our
attention to presequentiality relations, a simple class of
sequentiality relations, we were able to focus on the skeleton of the
definability preorder. The advantage of looking at presequentiality
relations is that we were able to completely characterize the set of
presequentiality relations under which a given function is invariant
via their p-level, a pair of integers which can be extracted from the
trace of the function.

We showed that interesting classes of functions
have natural characterizations in terms of p-levels, namely stable
functions, unstable functions, stable-dominating functions and
monovalued functions, and moreover exhibited natural hierarchies
within those classes of functions, hierarchies that make up
the skeleton of the definability preorder. We were also able to
completely charaterize various well-known functions in terms of
p-levels: any function with a p-level of $(2,2)$ is equiparallel to
\funbp, any function with a p-level of $(\infty,1)$ is equiparallel to 
\fundet, any function with a p-level of $(2,1)$ is equiparallel to
\funpor.

The keys to the p-level characterization are clearly the Reduction and 
Closure Lemmas, which allow us to derive canonical representatives 
for large classes of presequentiality relations. The characterization
itself is based on the fact that only two canonical presequentiality
relations are needed to describe the full set of presequentiality
relations under which a function is invariant. The next obvious step
in the investigation is to extend this result to full sequentiality
realtions. The question becomes: can we find canonical representatives 
of classes of sequentiality relations? A look at more
complicated examples of sequentiality relations (for example, the ones
used in the proof in \cite{Bucciarelli97}, or in the proof of the
strictness of the $\funbp+\por{i}$ hierarchy in Proposition
\ref{prop:6.4}) indicates that canonical representatives for full
sequentiality relations are far less nicely characterized than their presequentiality
counterparts. This is an area of future 
work, along the lines of the hypergraph approach of
\cite{Bucciarelli97,UNSTABLE:Bucciarelli97}. Another area of future
work is a study of unstable functions (or unstable 
degrees of parallelism). The structure of p-levels for stable
functions is richer than for unstable functions. Moreover,
Bucciarelli's original hierarchy fully lives in the \STABLE{}
semilattice. It would be interesting to see if the structure of the
\UNSTABLE{} semilattice is equivalently complicated, or simpler in some
respect. 

\textbf{Acknowledgments}. Thanks to the anonymous referees for
suggestions that helped improve and tighten the presentation, and for
various technical corrections.

\end{document}